# A method for simulating interfacial mass transfer on arbitrary meshes

G. Giustini*, R. I. Issa, M. J. Bluck

Nuclear Engineering Group, Mechanical Engineering Department

Imperial College London,  SW7 2AZ, UK

United Kingdom

* Corresponding author

g.giustini12@imperial.ac.uk; r.issa@imperial.ac.uk; m.bluck@imperial.ac.uk

**Abstract**

This paper presents a method for modelling interfacial mass transfer in Interface Capturing simulations of two-phase flow with phase change. The model enables mechanistic prediction of the local rate of phase change at the vapour-liquid interface on arbitrary computational meshes and is applicable to realistic cases involving two-phase mixtures with large density ratios. The simulation methodology is based on the Volume Of Fluid (VOF) representation of the flow, whereby an interfacial region in which mass transfer occurs is implicitly identified by a phase indicator, in this case the volume fraction of liquid, which varies from the value pertaining to the 'bulk' liquid to the value of the bulk vapour. The novel methodology proposed here has been implemented using the Finite Volume framework and solution methods typical of 'industrial' CFD practice. The proposed methodology for capturing mass transfer is applicable to arbitrary meshes without the need to introduce elaborate but artificial smearing of the mass transfer term as is often done in other techniques. The method has been validated via comparison with analytical solutions for planar interface evaporation and bubble growth test cases, and against experimental observations of steam bubble growth.

# 1 INTRODUCTION

Computation of two-phase flows with phase change requires combined solution of the hydrodynamics and heat transfer processes in the liquid and vapour phases, plus some form of tracking of the interface motion and of the local rate of interfacial mass transfer. A mathematical model that is general enough to be applied to any phase change process does not exist yet due to the multi-physics and multi-scale nature of the problem. Nevertheless, the development of Interface Capturing modelling techniques [1], such as the Volume Of Fluid (VOF) and Level Set methods, at the continuum scale enabled significant progress to be made in microscopic modelling of phase change phenomena in two-phase flow, at least in laboratory conditions. The vast majority of methods [2-4] for modelling mass transfer have been implemented in research codes applicable only to structured meshes and elementary geometries, and the only method [5] so far conceived to be applicable to arbitrary meshes and complex geometries typical of 'industrial' CFD is affected by serious shortcomings.

Here we present an Interface Capturing framework for simulation of multiphase flows with phase change on arbitrary meshes. The method's cornerstone is a straightforward representation of interfacial mass transfer that is applicable to any vapour-liquid phase change phenomenon and arbitrary fluids and is amenable to simulation on arbitrary meshes. For demonstration, the methodology was implemented using the open-source CFD toolbox OpenFOAM and was verified with a set of test cases that entails simulation of real fluids with large density difference and strong interfacial evaporation. The aim of this article is to demonstrate that the proposed method retains the capabilities of state-of-the-art methods for capturing mass transfer while extending their range of applicability to arbitrary meshes and eliminating the elaborate measures introduced by the Ref [5] methodology.

Ref [6] is considered to be the first to apply the VOF method to simulate two-phase flow with interfacial mass transfer using mechanistic evaluation of the rate of phase change at the vapour-liquid interface. Similar VOF-based methods were later adopted by a number of workers and applied to model nucleate boiling [7-9] [10], film boiling [5] and boiling in microchannels [11, 12]. The most popular VOF implementation, based on the work of Ref [5], has been implemented in the Computational Fluid Dynamics (CFD) codes ANSYS [12] and OpenFOAM [8], and is therefore in principle applicable to arbitrary geometries and computational meshes typical of industrial CFD simulation. However, that methodology presents serious shortcomings due to the introduction of numerical smearing of the interfacial rate of phase change, which is some distance removed from the true interfacial region, being artificially relocated in order to achieve numerical stability. Despite shortcomings introduced by such a spurious numerical 'cropping' of the physical rate of phase-change, which the current method eliminates, the work of Ref [5] is so far the only Interface Capturing method for phase change that has been implemented in industrial CFD codes. The Level-Set [3, 13] [14] and Front Tracking [2, 15] methods were also applied to various phase change problems, however these methods have not yet been used in industrial CFD practice and their application has been limited to idealised test cases; they also suffer from problems in conserving volume and mass. A different approach, based on the Constrained Interpolation Profile (CIP) method [16], has been implemented in a general-purpose two-phase flow solver [4] [17], which has been extensively used for modelling boiling phenomena [18]. The accuracy of the methodology was demonstrated both via comparison with analytical solutions and experimental data [19] and that method is therefore considered as a benchmark for numerical simulation of phase change in two-phase flow. Despite these superior attributes, the method is still not applicable on arbitrary meshes. Likewise, in Ref [20] a capable Phase Field model for quantitative prediction of mass transfer phenomena, albeit only applicable to elementary geometries, was developed. In all the above methods, a common approach to modelling phase-change at the vapour-liquid interface is based on explicit evaluation of the heat flux jump ('Stefan condition' [21] [1]) at the phase boundary, which requires explicit reconstruction of the interface and of the interfacial temperature gradients, typically resulting in unwieldy procedures [3, 4, 7] in order to extrapolate the temperature gradient from/to each side of the interface. An alternative approach has been derived from first principles within the framework of near-equilibrium thermodynamics and applied to Phase Field simulations of boiling phenomena [20]. The treatment of mass transfer in Ref [20] is based on the principles of equilibrium thermodynamics and was used to determine consistent formulations for the rate of phase change (as a function of the local deviations of interface temperature from its saturation value) and for the interfacial heat balance equation in near equilibrium conditions. An appropriate version of the general approach, suitable for incorporation into a VOF-based methodology, is that of Ref [22], in which the rate of phase change depends explicitly on the interface thickness, which cannot be justified on physical grounds. Different values of the interface thickness are used in the two implementations of the method reported so far [22] [20]; however, neither author provided justification for any particular imposed value of the interface thickness. Use of other methods, such as Lattice Boltzmann simulation [23] [24], has so far been limited to idealised fluids and test cases; also narrow is the range of applicability of conventional unstructured-mesh CFD approaches, which have typically been developed as application-specific tools [25, 26].

The majority of the Interface Capturing methods discussed above are embodied in research codes of limited industrial applicability; their implementation in the generally applicable framework of 'industrial' CFD for simulation on arbitrary geometries and computational meshes is still in the early stages of development. It is therefore highly desirable to generalise the phase-change modelling capabilities of current CFD approaches and enable their application to the simulation of two-phase flow on arbitrary meshes. The perspectives of such an enabling technology are applications to modelling boiling phenomena in industrial applications, requiring an ability to model mass transfer in geometrically complex flow configurations [27, 28] and to solve accurately the boiling model equations for real fluids on irregular grids, which are outside the current capabilities of CFD.

In this paper, we present the development of a methodology to capture mass transfer in Interface Capturing simulations of two-phase flows on arbitrary grids, using the CFD toolbox OpenFOAM [29]. The proposed simulation methodology is a novel approach that builds on the strengths of current general modelling techniques and extends their applicability to arbitrary meshes through implementation in a flow solver based on the VOF Interface Capturing method. The methodology is suitable for modelling boiling and bubble growth in industrially relevant fluids (e.g. water boiling at atmospheric pressure), which represent a real challenge for most Interface Capturing numerical methodologies, due to the large density difference between the liquid and vapour phases and the presence of strong evaporation. Therefore, a set of planar surface evaporation and bubble growth test cases has been chosen to test the proposed methodology. The objective of this paper is to demonstrate the accuracy of the current simulation methodology, via comparison with analytical solutions and laboratory observations of steam bubble growth in realistic boiling conditions.

## 2 MATHEMATICAL MODEL

The proposed simulation framework comprises a fluid flow model (Section 2.1), a method for interface capturing (Section 2.2) and a thermal model (Section 2.4). These are augmented in order to simulate, at the vapour-liquid interface, the effect of mass transfer due to phase change. The rate of phase change is computed mechanistically at the interface based on the predicted local temperature distribution (Section 2.3).

### 2.1 Fluid flow model

The incompressible Navier-Stokes equations can be written as

$$\frac{\partial \boldsymbol{u}}{\partial t} = -\nabla \cdot (\boldsymbol{u} \otimes \boldsymbol{u}) + \nabla \boldsymbol{T} - \frac{1}{\rho}\nabla p + \frac{1}{\rho}\boldsymbol{f} \qquad (1),$$

where $\rho$ is the density, $\boldsymbol{u}$ the fluid velocity, $\boldsymbol{T} = \frac{1}{2}\nu[\nabla \boldsymbol{u} + (\nabla \boldsymbol{u})^T]$ ($\nu$ being the kinematic viscosity), $p$ the pressure and $\boldsymbol{f} = \boldsymbol{f}_g + \boldsymbol{f}_\sigma$ the body force, which in this case includes a gravitational term $\boldsymbol{f}_g = \rho \boldsymbol{g}$ (being $\boldsymbol{g}$ the gravitational acceleration) and a suitable approximation $\boldsymbol{f}_\sigma$ of the surface tension force to be discussed in a later section.

For incompressible flows with phase change under investigation, the velocity divergence $\nabla \cdot \boldsymbol{u}$ is linked to the volumetric rate of phase change $\dot{m}\left[\frac{kg}{m^3 s}\right]$ through a continuity constraint (see Ref [4] for derivation):

$$\nabla \cdot \boldsymbol{u} = \dot{m}\left(\frac{1}{\rho_v} - \frac{1}{\rho_l}\right) \qquad (2).$$

In equation (2) and in the following, the subscript $l$ indicates the liquid phase and $v$ the vapour phase.

The current methodology is based on the Volume Of Fluid (VOF) method [1], whereby the phases are identified with the volume fraction of liquid $\alpha$, which is used to compute a generic (or mixture) fluid property, say the density $\rho$, as

$$\rho = \alpha \rho_l + (1-\alpha)\rho_v \qquad (3).$$

The phase indicator $\alpha$ varies from 1 (in the liquid) to 0 (in the vapour) across the fluid interface, whereby the phase indicator (in this case the liquid volume fraction) is used to model a transition [5] between the bulk phases, as opposed to other methods ([3, 4] [30]) that are based on reconstructing explicitly the geometry of the phase boundary.

### 2.2 Interface capturing model

It is assumed that the vapour and liquid phases are separated by a sharp interface. The interface is advanced through time using the model of Ref [20], which is applicable to conditions where thermally driven phase change occurs. The model equation for a generic conserved phase indicator derived in Ref [20] is used here to advance in time the liquid volume fraction distribution:

$$\frac{\partial \alpha}{\partial t} + \nabla \cdot (\alpha \boldsymbol{u}) = -\frac{\dot{m}}{\rho_l} + (2\alpha - 1)\nabla \cdot \boldsymbol{u} \qquad (4).$$

The last term in the equation is operative only in the interface region which is inevitably diffused numerically over a few mesh cells and it arises from the consideration that the liquid and vapour phases have different velocities within that region, as derived in Ref [20].

Using the continuity relation (2), the $\nabla \cdot \boldsymbol{u}$ term can be eliminated from the right-hand side of equation (4), which can be expressed in terms of the rate of phase change $\dot{m}$ and rewritten as

$$\frac{\partial \alpha}{\partial t} + \nabla \cdot (\alpha \boldsymbol{u}) = 2\dot{m}\left(\frac{1}{\rho_v} - \frac{1}{\rho_l}\right)\alpha - \frac{\dot{m}}{\rho_v} \qquad (5).$$

Ordinary interpolation schemes fail to maintain a constant interface thickness when used to discretize an advection equation such as (5) and typically cause unphysical diffusion of the interfacial region. The problem of interface diffusion is well known in computational science and a popular method to alleviate it has been proposed by Ref [31], who introduced a modified advection equation, augmented by additional compressive terms for maintaining the sharp interfaces. Equation (5) is therefore augmented to include interface 'sharpening' terms

$$\frac{\partial \alpha}{\partial t} + \nabla \cdot (\alpha \boldsymbol{u}) + \nabla \cdot [\alpha(1-\alpha)\boldsymbol{m}] = \nabla \cdot (\varepsilon \nabla \alpha) + 2\dot{m}\left(\frac{1}{\rho_v} - \frac{1}{\rho_l}\right)\alpha - \frac{\dot{m}}{\rho_v} \qquad (6),$$

where the $\nabla \cdot [\alpha(1-\alpha)\boldsymbol{m}]$ term promotes interface compression and $\nabla \cdot (\varepsilon \nabla \alpha)$ is a diffusive term. Here $\boldsymbol{m} = \frac{\nabla \alpha}{|\nabla \alpha|}$ is the interface unit normal vector and $\varepsilon$ is the desired interface thickness, which needs to be specified as an input to the model. Proof of the effect of the additional terms in equation (6) was illustrated from first principles in the original work [31] in a Level-Set framework (in the absence of phase change); the Ref [31] sharpening technique was also successfully applied to flows with phase change in Ref [4], using the Constrained Interpolation Profile [16] method for advancing the interface. In this work, as in the previous studies that used the Ref [31] sharpening method, the interface thickness $\varepsilon$ is related to the grid size.

The interface transport model here proposed has been solved with the standard MULES solver available in the OpenFOAM code [32].

## 2.3 Calculation of the rate of phase change

The rate of phase change is computed at the interface from the local temperature distribution. Various approaches can be used to this end.

A straightforward method [5] is based on considerations at the molecular level [33] and uses a basic kinetic theory description of the interfacial phase change process to link local deviations of the interface temperature from its saturation value to the rate of phase change pertaining to the continuum fluid model. With the kinetic approach, the rate of phase change can be computed in the interfacial region as

$$\dot{m} = \varphi(T - T_{SAT})|\nabla \alpha| \qquad (7),$$

where $T_{SAT}$ is taken as the saturation temperature at the externally imposed pressure and the coefficient $\varphi$ is computed as

$$\varphi = \frac{2\hat{\sigma}}{2-\hat{\sigma}} \left(\frac{M}{2\pi R_g}\right)^{1/2} \frac{\rho_v h_{lv}}{T_{SAT}^{3/2}} \qquad (8),$$

where $h_{lv}$ is the latent heat of vaporization, $M$ the molar mass and $R_g$ the universal gas constant. The assumption implicit in equation (7) is that the portion of surface area of vapour-liquid interface contained in one cell can be expressed as $S_{INT} = |\nabla\alpha|V_{cell}$. $\dot{m} \neq 0$ only in the interfacial region where $|\nabla\alpha| \neq 0$ and $\dot{m} = 0$ elsewhere; the rate of phase change is positive for the case of evaporation ($T > T_{SAT}$) and negative for the case of condensation ($T < T_{SAT}$). A theory for predicting the 'accommodation' [34] coefficient $\hat{\sigma}$ does not yet exist, and therefore the appropriate value for $\hat{\sigma}$ needs to be extracted from experiment. Various experimental sources indicate a value close to 1, as summarised in a review [34] that covers experimental works on evaporation and condensation [35] spanning several decades. Here a value of 1 is retained for the coefficient $\hat{\sigma}$. It was found in [5, 11] that flow solvers using equation (7) for modelling mass transfer were subject to numerical instabilities in the case of strong evaporation and large density ratios $\rho_l/\rho_v$, as is the case in most flows of practical interest. In order to alleviate the issue, a 'cropping' procedure was introduced, whereby the source terms due to mass transfer are artificially removed from the interfacial region and located some distance away. The current solver is immune to such issues and it has therefore been possible to eliminate the unphysical 'cropping' step.

For all simulations presented in this paper, the volumetric rate of phase change $\dot{m}$ is computed with equation (7).

**2.4 Thermal model**

In most flows of practical interest, interfacial mass transfer is driven by diffusion of heat towards or away from the vapour-liquid interface; in typical boiling conditions, for example, evaporation is driven by the flow of heat from the liquid towards the interface. Therefore, it is necessary to solve an appropriate enthalpy balance equation in order to obtain the temperature field required for evaluation of the rate of phase change. To this end, a single-fluid enthalpy is defined as

$$h = \frac{[\alpha\rho_l c_l + (1-\alpha)\rho_v c_v](T - T_{REF})}{\alpha\rho_l + (1-\alpha)\rho_v} \qquad (9),$$

where $c$ is the specific heat capacity and a possible choice of the reference temperature is the saturation temperature at the prevailing system pressure, that is, $T_{REF} = T_{SAT}$.

Enthalpy transport is modelled with the following balance equation

$$\frac{\partial \rho h}{\partial t} + \nabla \cdot (\rho \boldsymbol{u} h) - \nabla \cdot (\rho D \nabla h) = S_h \qquad (10),$$

where $D = \frac{k}{\rho c}$ is the mixture thermal diffusivity.

The volumetric heat sink $S_h$ is due to phase change and is modelled after Refs [20, 22] as

$$S_h = -\dot{m}\left[h_{lv} + (\rho_l c_l - \rho_v c_v)\frac{\varepsilon}{\sqrt{2}}\left(\frac{1}{\rho_v} - \frac{1}{\rho_l}\right)\boldsymbol{m} \cdot \nabla T\right] \qquad (11),$$

where, as noted in the derivation presented in Ref [20], $\varepsilon$ is the same quantity appearing also in the interface advection equation (6). Equation (10) can be solved with any standard iterative method. In the current simulations, use of a pre-conditioned bi-conjugate gradient solver, available in OpenFOAM [29], was made. Note that owing to the current formulation of mass transfer it is possible to model the effect of phase change on the energy balance with a volumetric sink term, which enables one to dispense with explicit reconstruction of the interfacial heat flux jump.

## 3 MODEL SOLUTION

### 3.1 Advecting the interfacial region

The interface is advanced through time using the standard MULES solver available in the OpenFOAM code. In semi-discrete form, equation (6) can be written for a control volume $O$ as

$$\frac{V_O}{\delta t}(\alpha_O^{NEW} - \alpha_O) + \sum_f \{\mathcal{L}(\alpha_O)\phi_f + \mathcal{L}[\alpha_O(1-\alpha_O)]\mathbf{m}_f \cdot \mathbf{S}_f\} = V_O S_\alpha \qquad (12)$$

Which is solved explicitly using the MULES solver

$$\alpha_O^{NEW} = \alpha_O - \frac{\delta t}{V_O} \sum_f \{\mathcal{L}(\alpha_O)\phi_f + \mathcal{L}[\alpha_O(1-\alpha_O)]\mathbf{m}_f \cdot \mathbf{S}_f\} + \delta t S_\alpha \qquad (13).$$

In the above equations, $V_O$ is the control volume, $f$ indicates the surfaces separating the control volume $O$ from each one of its neighbours $i$, $\delta t$ the time step, $\mathcal{L}$ indicates linear interpolation from cell centres to cell faces, $\phi_f$ is the face flux computed from the nodal velocity $\mathbf{u}_O$ as $\phi_f = \mathcal{L}(\mathbf{u}_O) \cdot \mathbf{S}_f$, $\mathbf{S}_f$ is the face area vector (normal to face $f$), and $\mathbf{m}_f$ the unit vector normal to the liquid-vapour interface evaluated at the centroid of face $f$. The terms on the right hand side of equation (6) are included in the source term $S_\alpha = \frac{\varepsilon}{V_O} \sum_f [\|\mathbf{S}_f\|(\nabla^\perp \alpha)_f] + \left[2\dot{m}_O\left(\frac{1}{\rho_v} - \frac{1}{\rho_l}\right)\alpha_O - \frac{\dot{m}_O}{\rho_v}\right]$, where $(\nabla^\perp \alpha)_f$ is the component of the volume fraction gradient normal to surface $f$.

### 3.1 Pressure-velocity coupling

Coupling between pressure and velocity is achieved with a PISO-based [36] [37] segregated algorithm.

With $O$ as the control volume under consideration and $i$ indicating neighbouring control volumes, the advection-diffusion matrix is unpacked into its diagonal and off-diagonal components:

$$\int_{V_O} [-\nabla \cdot (\mathbf{u} \otimes \mathbf{u}) + \nabla T] dV \rightarrow \mathbf{H}(\mathbf{u}) - A_O \mathbf{u}_O \qquad (14),$$

where

$$\mathbf{H}(\mathbf{u}) = \sum_i A_i \mathbf{u}_i, \quad A_O = \sum_i A_i \qquad (15),$$

and the details of the particular interpolation schemes used to generate the coefficients $A_i$ are specified in a later section. The central coefficient $A_O$ is augmented by a contribution $A_O^0 = \frac{V_O}{\delta t}$ due to the temporal derivative. The form of the central coefficient $A$ so computed, which includes $A_O^0$, depends on the chosen temporal discretization scheme, also to be specified in a later section.

As is common in modelling buoyant flows, the hydrostatic pressure is lumped together with the static pressure to give the piezometric pressure as a main working variable $p_z = p - \rho \mathbf{g} \cdot \mathbf{x}$ with the body force being computed as $\mathbf{f}_g = -(\mathbf{g} \cdot \mathbf{x})\nabla \rho$, $\mathbf{x}$ being the position vector.

In the adopted segregated solution procedure, the nodal velocity $\boldsymbol{u}_O$ is advanced in time and corrected as follows: First, the component of the velocity normal to face $f$ (or *flux*) is estimated as

$$\phi_f = \mathcal{L}\left(\frac{H(\boldsymbol{u})}{A}\right) \cdot \boldsymbol{S}_f + \frac{\mathcal{L}(f) \cdot \boldsymbol{S}_f}{\bar{\rho}_f \mathcal{L}(A)} \qquad (16),$$

with the face density $\bar{\rho}_f$ taken as the harmonic average [38] of densities in the two adjacent cells:

$$\frac{1}{\bar{\rho}_f} = \frac{w_O}{\rho_O} + \frac{w_i}{\rho_i} \qquad (17),$$

where $w_O$ and $w_i$ interpolation factors based on the distance of each cell centre $i$ or $O$ from the face centre $f$. For a uniform grid, the face density would be

$$\frac{1}{\bar{\rho}_f} = \frac{1}{2}\left(\frac{1}{\rho_O} + \frac{1}{\rho_i}\right) \rightarrow \bar{\rho}_f = \frac{2\rho_O\rho_i}{\rho_O + \rho_i} \qquad (18).$$

Here $f$ indicates the surface separating the control volume $O$ from each one of its neighbours $i$, $\mathcal{L}$ indicates linear interpolation from cell centres to cell faces, $\boldsymbol{S}_f$ is the face area vector (normal to face $f$). The surface tension force $\boldsymbol{f}_\sigma$ is computed from the interface curvature using the well-established Continuum Surface Force (CSF) method of Ref [39] as

$$\boldsymbol{f}_\sigma = \sigma \kappa \nabla \alpha \qquad (19),$$

where $\sigma$ is the surface tension coefficient and the curvature $\kappa$ is computed as

$$\kappa = -\nabla \cdot \boldsymbol{m} \qquad (20).$$

The pressure field is computed by solving a modified equation that includes the source term $\dot{m}\left(\frac{1}{\rho_v} - \frac{1}{\rho_l}\right)$ due to mass transfer, that is:

$$\sum_f \frac{S_f}{\bar{\rho}_f \mathcal{L}(A)} \Delta_f p_z = \sum_f \phi_f - \dot{m}\left(\frac{1}{\rho_v} - \frac{1}{\rho_l}\right) \qquad (21),$$

where $\Delta_f p_z$ is the piezometric pressure difference across face $f$. The flux is then corrected using the newly calculated pressure field, and used to update the nodal velocity as

$$\boldsymbol{u}_O^{n+1} = \boldsymbol{u}_O^n + \frac{1}{\rho A} \mathcal{R}\left\{\frac{\left[\phi_f^{n+1} - \mathcal{L}\left(\frac{H(\boldsymbol{u})}{A}\right) \cdot \boldsymbol{S}_f\right]}{\bar{\rho}_f \mathcal{L}(A)}\right\}$$

$$(22),$$

$n$ being an integer used to count corrector steps.

The corrected velocity can be used to recalculate $H(\boldsymbol{u})$ and repeat the corrector step. The reconstruction operator [40] [41] is defined as $\mathcal{R}(a_f) = \left(\sum_f \frac{\boldsymbol{S}_f \otimes \boldsymbol{S}_f}{S_f}\right)^{-1} \cdot \left[\sum_f a_f \frac{\boldsymbol{S}_f}{S_f}\right]$, $a_f$ being any scalar defined at cell faces and $S_f = \|\boldsymbol{S}_f\|$.

For unsteady simulations, two corrector steps are sufficient for convergence within a time step [36, 42]. The sequence of steps for enforcing coupling of the pressure and velocity fields highlighted in the preceding text is in the same spirit as that which may be employed by any standard segregated flow solver; the modifications introduced here are for extending the solution procedure to two-phase flows (equation 16) and for including the effect of mass transfer (equation 21).

**3.2 Solution loop**

The solution procedure entails solving the Navier-Stokes equations, including the volumetric CSF approximation of the surface tension force and the body force, the energy and interface advection equations, and evaluating the mass transfer model.

For every time step, the segregated solution sequence is as follows:

1. Evaluate equation (13) to advance the interface using a known rate of phase change from the previous time step.
2. Update the single fluid properties (equation (3)).
3. Compute the surface tension force with the CSF method.
4. Compute the rate of phase change with equation (7).
5. Solve the enthalpy equation (10).
6. Recalculate the rate of phase change.
7. Compute the velocity and pressure fields via the desired number of PISO steps (as outlined in Section 3.1).

The Euler scheme was used for temporal discretization, therefore in the discretized equations $A = A_O + A_O^0$. For the current simulations, two 'corrector' steps were used in the PISO loop. The time step was dynamically updated based on the CFL condition, $\delta t \leq S_{CFL} \frac{\Delta x}{|\boldsymbol{u}|}$, where $\Delta x$ is the smallest cell size, $|\boldsymbol{u}|$ is the maximum value of the velocity magnitude in the domain (enabling dynamic update of the time step during the simulation), and $S_{CFL}$ is a safety factor here set equal to 0.2. In conditions of interest, typical flow velocities driven by mass transfer are of the order 0.1 – 1.0 m/s, which correspond, for cell sizes as small as 0.8 micron typical of the current simulations, to time step values around $10^{-7}$ s.

All numerical schemes used for discretization of the model equations have been selected from the standard choices available in the OpenFOAM code and are summarized in Table 1. Unless specified otherwise in Table 1, linear interpolation has been used for divergence terms and the Gauss theorem for gradient terms. Further details on OpenFOAM discretization schemes are available online [29] and in the Ref [32] thesis.

| Term | Scheme |
|---|---|
| $\nabla \cdot (\rho \boldsymbol{uu})$ | Van Leer |
| $\nabla \cdot (\alpha \boldsymbol{u})$ | Van Leer |
| $\nabla \cdot [\alpha(1-\alpha)\boldsymbol{m}]$ | 'Interface compression' [32] |
| $\nabla \cdot (\rho \boldsymbol{u} h)$ | Van Leer |

Table 1

Discretization schemes employed. If not specified in the table, linear interpolation and the Gauss theorem have been used for discretization of, respectively, divergence and gradient terms.

## 4 VALIDATION - PLANAR INTERFACE TEST CASES

In order to validate the current method, simple one-dimensional planar interface test cases are considered first.

In the Stefan problem test (Section 4.1), the evaporation-driven motion of a flat interface separating two fluids with equal density is considered; for this test, owing to the equal density assumption, there is no fluid dilatation due to evaporation, the domain boundaries are impermeable and the interface moves solely due to the local effect of mass transfer.

In the sucking interface problem (Section 4.2), another one-dimensional flat-interface benchmark, saturated water at atmospheric pressure is the test fluid, with a density ratio $\frac{\rho_l}{\rho_v} \cong$ 1600. In this case, the domain boundary on the liquid side is open and the interface moves due to the pushing of the interface caused by vapour generation.

### 4.1 Stefan problem

The Stefan problem concerns the prediction of the evaporation-driven motion of a planar interface due to transport of heat from superheated vapour. A fictitious fluid is considered for which the two phases have equal density. The boundaries of the domain, sketched in Figure 1, are impermeable, no flow is allowed into or out of the domain and the interface moves only due to mass transfer. The one-dimensional domain is 1 mm long and the temperature of the heated wall $T_w$ is 383.15K. The opposite wall is a at temperature of 373.15K which is also taken as the saturation temperature of the fictitious fluid. Interface position and temperature distribution are known from theory [5] and can be expressed as

$$x_i(t) = 2\zeta\sqrt{D_v t} \qquad (23) \text{ (interface position)}$$

$$T(x,t) = T_w - \frac{(T_w - T_{SAT})}{\text{erf}(\zeta)} \text{erf}\left(\frac{x}{2\sqrt{D_v t}}\right) \qquad (24) \text{ (temperature distribution),}$$

where $D_v = \frac{k_v}{\rho_v c_v}$ is the vapour thermal diffusivity and the constant $\zeta$ is computed from the solution of the following equation

$$\zeta \exp(\zeta^2) \text{erf}(\zeta) = \frac{c_v(T_w - T_{SAT})}{\sqrt{\pi} h_{lv}} \qquad (25).$$

The fluid properties listed in Table 2 have been used to evaluate the above expressions. For uniform grids considered here and in the following, $\varepsilon = \frac{\Delta x}{2}$, as in previous applications of the Ref [31] interface sharpening method reported above. The initial condition of the simulation corresponds to the analytical solution after 0.03 s and the simulation is continued until 0.12 s.

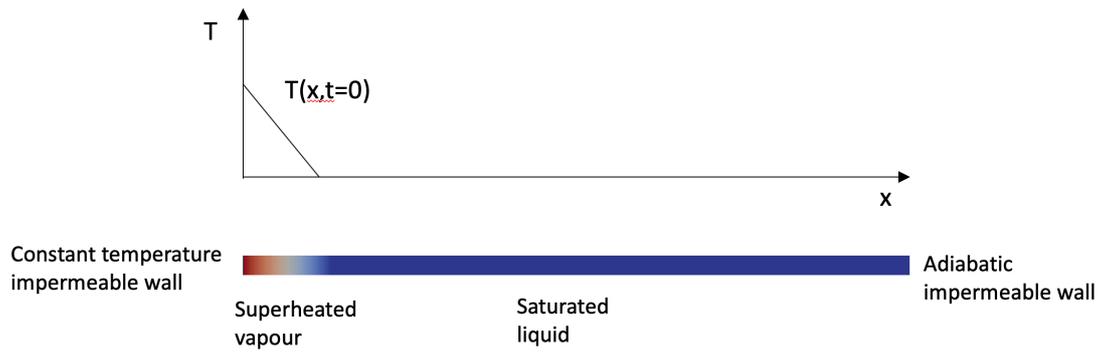

Figure 1

Simulation setup for a planar interface evaporation test case with impermeable domain boundaries, showing the initial temperature distribution (blue-red color scale) from analytical solution.

| Fictitious fluid properties | | |
|---|---|---|
| **Property** | **Vapour** | **Liquid** |
| Dynamic viscosity $\mu$ [$Pa \cdot s$] | $1.0 \times 10^{-5}$ | 0.01 |
| Density $\rho$ [$kg/m^3$] | 1.0 | 1.0 |
| Specific heat capacity $c$ [$J/Kg/K$] | 1000.0 | 1000.0 |
| Thermal conductivity $k$ [$W/m/K$] | 0.01 | 1.0 |
| Surface tension coefficient $\sigma$ [$N/m$] | 0.01 | |
| Latent heat of vaporization $h_{lv}$ [$J/Kg$] | $1000.0 \times 10^3$ | |
| Gas constant $R = R_g/M$ [$J/Kg/K$] | 461.52 | |

Table 2

Physical properties of the fictitious fluid considered for the Stefan problem.

As shown in Figure 2, the time history of the interface position matches the analytical solution, confirming the ability of the current methodology to capture interface motion due to interfacial mass transfer for the case of fluids with equal density.

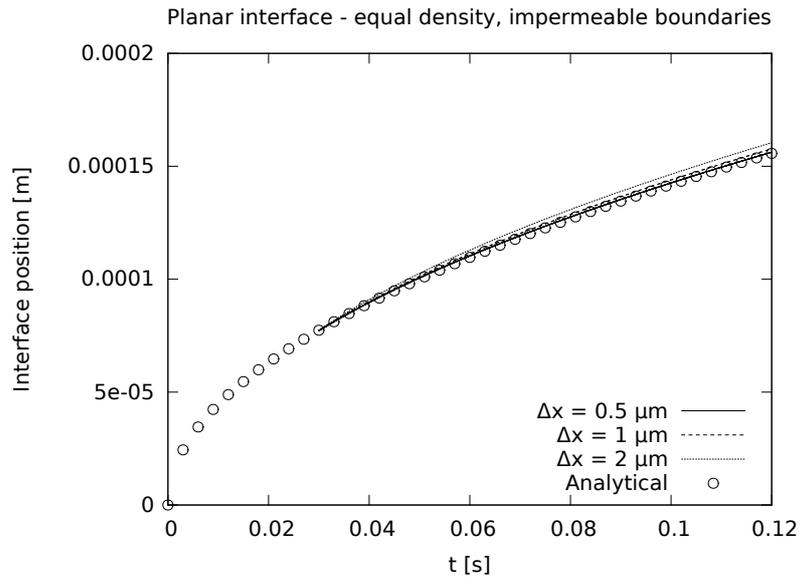

Figure 2

Interface position versus time for the case of mass transfer driven motion of a planar interface separating two fluids with equal density.

In the following sections, additional effects of large fluid density ratio are introduced. The main challenge posed by the density difference is that the generation of vapour causes the fluid volume to increase, which needs to be accommodated by a displacement of the interface. For incompressible fluids, this can happen only if at least one of the domain boundaries are open to allow outflow, as is the case of all the tests that follow.

**4.2  Sucking interface problem**

The physical problem considered here is that of predicting the motion of a planar interface, separating liquid and vapour phases, due to the transport of heat from the remote superheated liquid (at temperature $T_\infty$) towards the interface. The vapour phase is assumed as saturated and confined by an adiabatic wall, whereas there is an open boundary on the liquid side of the one-dimensional simulation domain sketched in Figure 3. Vapour is generated at the liquid-vapour interface, which causes the interface to be pushed towards the open boundary of the domain.

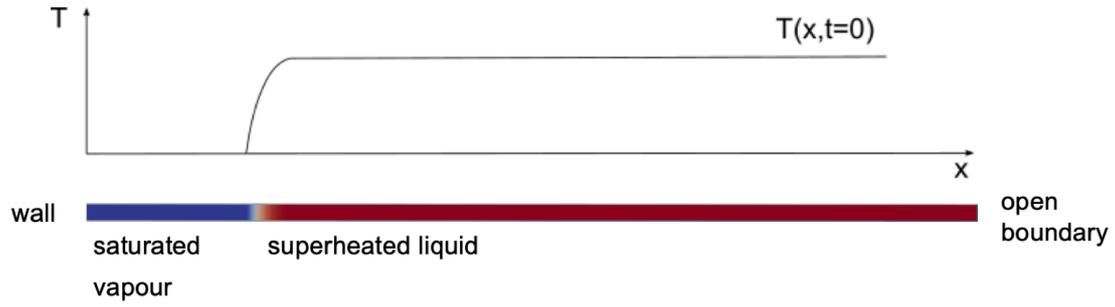

Figure 3

Simulation setup for the planar interface evaporation test case, showing the initial temperature distribution from analytical solution.

For this case of thermally-driven motion of a planar interface, closed-form expressions were derived in Ref [6] for the temporal variation of the interface position and of the temperature distribution in the liquid. The interface position varies with time as

$$x_i(t) = 2\eta\sqrt{D_v t} \qquad (26),$$

the constant $\eta$ is computed via solution of the following equation

$$\eta - \frac{f(\eta)}{g(\eta)} = 0 \qquad (27),$$

with

$$f(\eta) = (T_\infty - T_{SAT})c_v k_v \sqrt{D_v} \exp\left(-\eta^2 \frac{\rho_v^2}{\rho_l^2}\frac{D_v}{D_l}\right) \qquad (28)$$

and

$$g(\eta) = h_{lv} k_v \sqrt{\pi D_l}\,\text{erfc}\left(\eta\frac{\rho_v}{\rho_l}\sqrt{\frac{D_v}{D_l}}\right) \qquad (29).$$

The corresponding temperature distribution in the liquid is

$$T(x,t) = T_\infty - (T_\infty - T_{SAT})\frac{\text{erfc}\left(\frac{x}{2\sqrt{2D_l t}}\right)}{\text{erfc}\left(\eta\frac{\rho_v}{\rho_l}\sqrt{\frac{D_v}{D_l}}\right)} \qquad (30).$$

For evaluation of the analytical solutions, the thermophysical properties of saturated water at 1 bar, as indicated in Table 3, have been used. $T_\infty$ is set equal to 383.15K.

| Properties of saturated water at 1 bar | | |
|---|---|---|
| Property | Vapour | Liquid |
| Dynamic viscosity $\mu$ [$Pa \cdot s$] | $12.2 \times 10^{-6}$ | $281.6 \times 10^{-6}$ |
| Density $\rho$ [$kg/m^3$] | 0.6 | 958.4 |
| Specific heat capacity $c$ [$J/Kg/K$] | 2077.5 | 4216.6 |

| Thermal conductivity $k$ [W/m/K] | $24.8 \times 10^{-3}$ | $677.8 \times 10^{-3}$ |
|---|---|---|
| Surface tension coefficient $\sigma$ [N/m] | 0.059 | |
| Latent heat of vaporization $h_{lv}$ [J/Kg] | $2258.0 \times 10^3$ | |
| Gas constant $R = R_g/M$ [J/Kg/K] | 461.52 | |

Table 3

Physical properties of saturated water at 1 bar.

In the simulation, the initial temperature distribution and interface position correspond to the analytical solution after 0.1 seconds; the simulation is continued for 1.1 seconds.

Mesh-convergent behaviour is observed in Figure 4, where the time history of interface position and the temperature distribution in the liquid at 0.65 seconds into the simulation are shown.

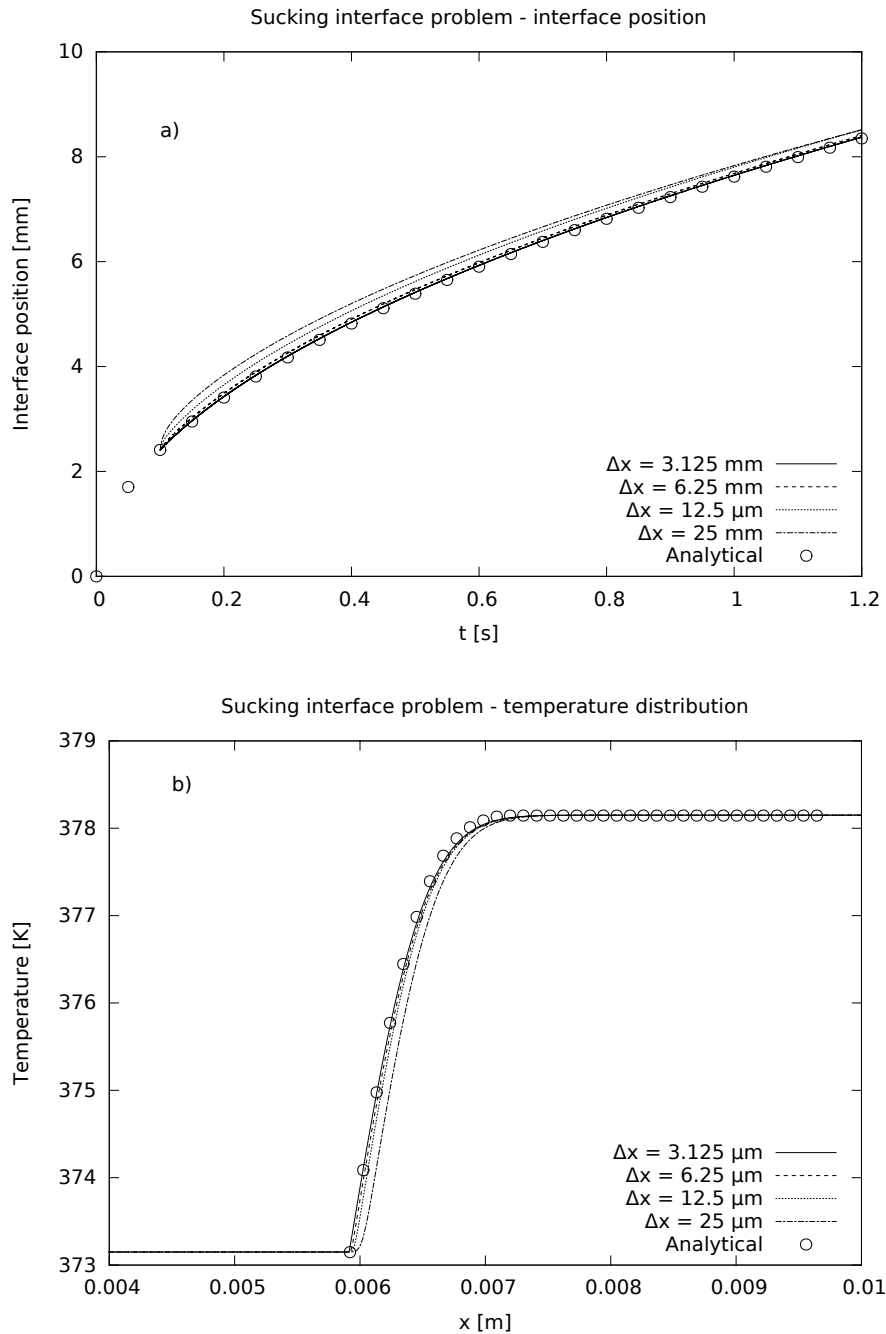

Figure 4

a) interface position versus time for the case of evaporation-driven motion of a planar interface, b) temperature distribution in the liquid at t = 0.65 s.

## 5 VALIDATION - ISOLATED BUBBLE GROWTH IN A POOL OF LIQUID

A set of isolated bubble growth problems is used to validate predictions of mass transfer at a curved interface.

The physical problem is that of predicting the growth of a spherical bubble in a uniformly superheated pool of liquid, away from solid surfaces and in the absence of gravity, as originally considered by the theory of Scriven presented in Ref [43]. It should be pointed out that as in the theoretical model, all computations presented here commence with an already formed bubble. These 'ideal' conditions are modelled in Section 5.1.2 and 5.1.4 for the cases of water boiling and sodium boiling. The additional sodium boiling test was chosen because the boiling behaviour of that fluid has not yet been adequately studied from a computational perspective; nevertheless, sodium is an important substance widely used in the energy conversion industry. The main difference between sodium and fluids, such as water or refrigerants, more commonly studied using CFD is that liquid sodium has a much higher thermal conductivity than water or any refrigerant.

Using a Scriven test for the case of water, the current method is validated on irregular polyhedral grids in Section 5.1.3. Such a qualitative test, using low-resolution grids to limit the computational effort, is used to demonstrate applicability of the method to arbitrary meshes.

Finally, the current method is validated in Section 5.2 and Section 6 against laboratory observations [44] [45] of steam bubble growth at a distance from walls in a uniformly heated pool of liquid.

In the test cases that follow, axisymmetric simulation has been used to compare results with spherical analytical solutions and experimental observations.

**5.1 Comparison with analytical solution**

**5.1.1 Analytical solution**

The growth of an isolated bubble in a uniformly superheated pool of liquid is one of the few boiling test cases for which a solution for the bubble radius as a function of time can be derived analytically [43]. The theory[1] assumes that a bubble grows due to the diffusion of heat from the remote liquid towards the curved surface of the bubble [46]. Under the assumption that the vapour pressure inside the bubble is the same as the remote liquid pressure (i.e. the externally imposed system pressure), it is assumed that the body of vapour in the bubble is at the saturation temperature corresponding to the externally imposed pressure, and that saturated vapour is generated at the bubble surface. Neglecting the effect of gravity and surface tension, the growth of the bubble is limited purely by the rate of diffusion of heat towards the bubble surface at $T_{SAT}$, from the remote liquid assumed at some higher temperature $T_\infty$. Scriven derived expressions for the bubble radius as a function of time and for the temperature distribution in the liquid as a function of time and of the radial distance $r$ from the bubble surface. These formulae are reported below and have been used for initialising the isolated bubble growth simulations. The Scriven theory predicts that the bubble radius increases proportionally to the square root of time through a growth constant $\beta$:

---

[1] The theory is predicated on the assumption that bubble growth is driven purely by thermal effects. Ref [37] showed that assuming thermally-driven bubble growth is appropriate at times after bubble inception larger than a characteristic time scale computed as $\tau = \frac{B^2}{A^2}$, where $B^2 = \frac{12}{\pi} D_l Ja$, $A^2 = \frac{2}{3} \frac{h_{lv} \rho_v c_v}{\rho_l T_{SAT}}$, the Jakob number is computed as $Ja = \frac{\Delta T \rho_l c_l}{\rho_v h_{lv}}$ and $\Delta T$ is the liquid superheat in excess of the saturation temperature $T_{SAT}$. All calculations presented in this work start at a time into bubble growth sufficiently large compared to the time constant $\tau$.

$$R(t) = 2\beta\sqrt{D_l t} \quad \text{(bubble radius)} \quad (31).$$

The growth constant is evaluated via solution of the following transcendental equation (a discussion of the applicability of the Scriven approach to computing the growth constant of bubbles is presented in [47]):

$$\frac{\rho_l c_l (T_\infty - T_{SAT})}{\rho_v (h_{lv} + (c_l - c_v)(T_\infty - T_{SAT}))} = 2\beta^2 \int_0^1 \exp\left(-\beta^2 \left((1-\xi)^{-2} - 2\left(1 - \frac{\rho_v}{\rho_l}\right)\xi - 1\right)\right) d\xi \quad \text{(growth constant equation)} \quad (32)$$

Once the growth constant and the temporal variation of bubble radius are known, the radial temperature distribution in the liquid (as noted, the vapour is assumed at the saturation temperature) can be computed at any time $t$ via evaluation of the following expression:

$$T(r,t) = T_\infty - 2\beta^2 \left(\frac{\rho_v(h_{lv} + (c_l - c_v)(T_\infty - T_{SAT}))}{\rho_l c_l}\right) \int_{1-\frac{R(t)}{r}}^1 \exp\left(-\beta^2 \left((1-\xi)^{-2} - 2\left(1 - \frac{\rho_v}{\rho_l}\right)\xi - 1\right)\right) d\xi$$
(temperature distribution in the liquid) (33).

An estimate of the approximate thickness $d_T$ of the thermal boundary layer in the liquid around the bubble can be computed from an energy balance at the bubble surface. The heat flux into the bubble can be computed at the bubble surface (vapour-liquid interface) as $\dot{q}''_{cond} = k_l \left(\frac{\partial T}{\partial r}\right)_{int} \cong \frac{k_l(T_\infty - T_{SAT})}{d_T}$, and the latent heat transfer due to evaporation can be expressed as $\dot{q}''_{lat} = \rho_v h_{lv} \left(\frac{\partial R}{\partial t}\right) = \rho_v h_{lv} \left(\frac{2\beta^2 D_l}{R(t)}\right)$. $\dot{q}''_{cond}$ and $\dot{q}''_{lat}$ must be equal and therefore $d_T \cong \frac{\rho_l c_l (T_\infty - T_{SAT})}{\rho_v 2\beta^2 h_{lv}} R(t)$.

The above set of expression can be used to compute bubble radius, temperature distribution and thermal boundary layer thickness at any time and is used in the following to compute the initial conditions of the simulations presented.

**5.1.2 Modelled bubble growth vs analytical solution - water**

Fluid properties are those of saturated water at atmospheric pressure indicated in Table 3. Initial conditions for the simulation are generated with the Scriven model equations (described in the preceding text) at $t_0 = 0.17\ ms$ into bubble growth (sufficiently large compared to $\tau \approx 7 \times 10^{-3}\ ms$). This is an arbitrary time chosen in order to obtain initial radius $R_0$ and initial thickness $d_T$ of the thermal boundary layer respectively equal to 0.11 mm and approximately 5 micron, values that are representative of the early stages of bubble growth in typical low-superheat laboratory conditions. The superheat of the remote liquid is equal to 3.1K, a value typical of laboratory experiments [44], and the corresponding bubble growth constant $\beta$, computed with equation (31) is equal to 9.4. The simulation follows the growth of the bubble until the bubble radius is approximately four times the initial radius. The axisymmetric computational domain is a square of $800 \times 800\ \mu m^2$ and is discretised with increasingly small uniform cells of 4.8, 3.2, 1.6 and 0.8 $\mu m$ in size. A diagram of the initial conditions is shown in Figure 5. In Figure 5a, the left boundary is the axis of symmetry, the bottom boundary is a symmetry plane, and the other two boundaries are open to the flow.

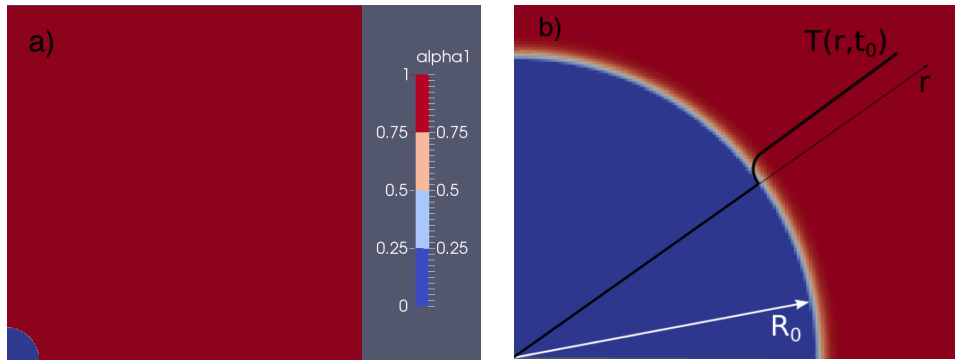

Figure 5

a) initial bubble shape and b) temperature distribution for the isolated bubble test case.

This test case is more demanding, from the point of view of numerical resolution, than the planar interface case (Section 4.2), as the grid size needs to be fine enough to resolve the very thin initial thermal boundary layer on the liquid side of the interface, which is initially about 5 microns thick. In practice it was found that only a few mesh points are required to resolve the initial thermal boundary layer, as indicated in Figure 6, where good agreement with analytical solution is observed for the two finest meshes that correspond, respectively, to 3 and 6 grid points in the initial thermal boundary layer. The spherical bubble tests were affected to some extent by the build-up of spurious currents, a well-known numerical artefact discussed, for example, in Refs [4] [38] [48]. The currents are due to irreducible errors inherent (according to Ref [39]) in the surface tension model, however they were of small enough magnitude for the numerical solutions not to be affected significantly.

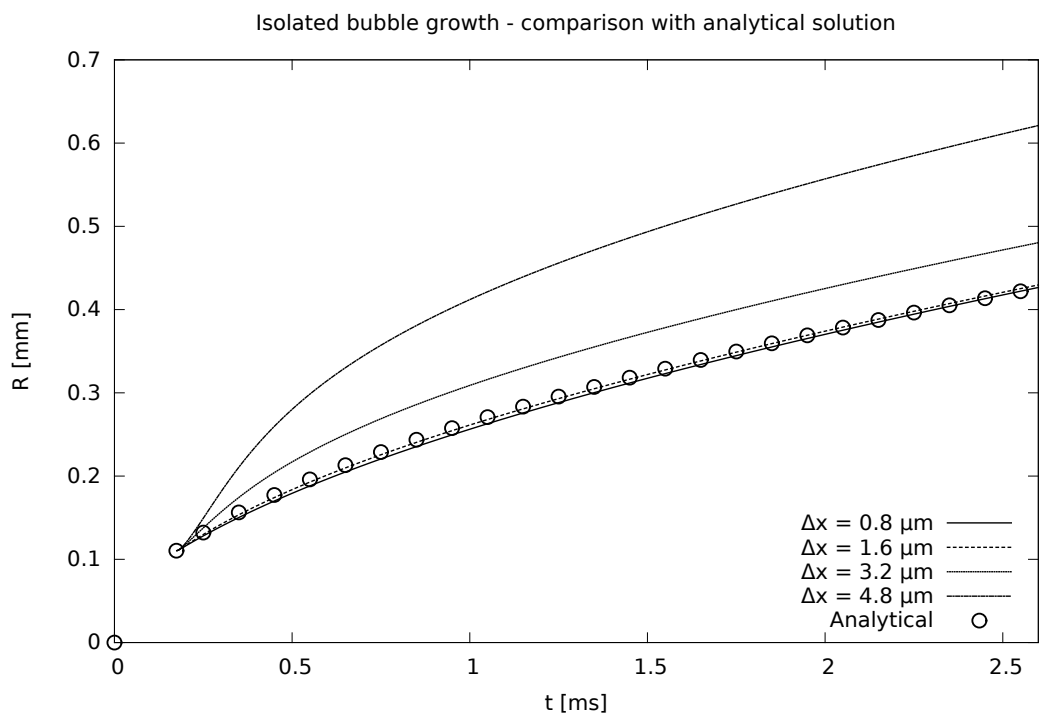

Figure 6

Comparison with analytical solution of modelled time history of bubble radius for the case of bubble growth in isolation in a uniformly superheated pool of liquid water.

**5.1.3 Simulation using an irregular polyhedral mesh**

The above test case was repeated using an irregular two-dimensional translationally symmetric grid, consisting of polyhedral cells of average size equal to approximately $\overline{\Delta x} \approx 4.5\ \mu m$, shown in Figure 7, where the initial volume fraction distribution is also indicated. For this test, $\varepsilon = \overline{\Delta x}/2$ and the results are compared with the case using a regular two-dimensional translationally symmetric grid with $\Delta x = 4.8\ \mu m$.

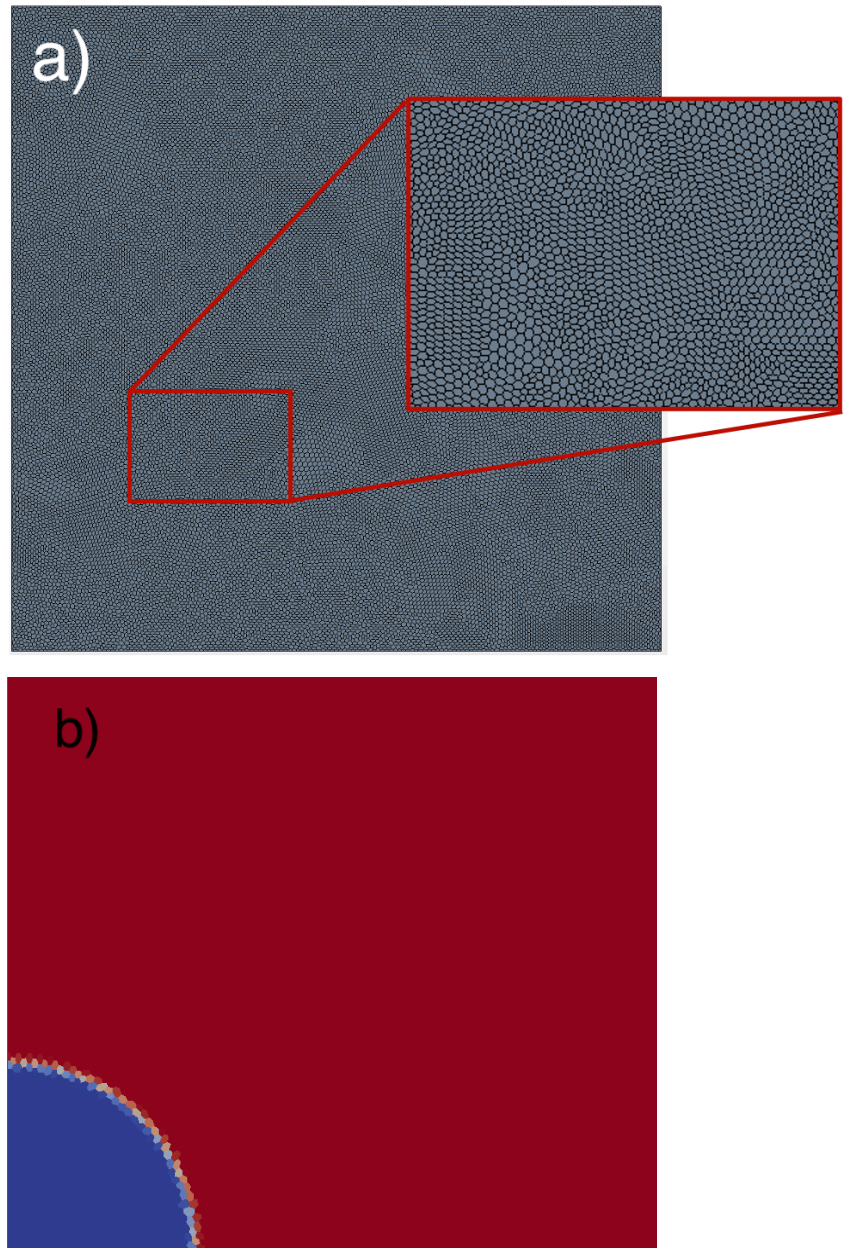

Figure 7

a) Polyhedral mesh and b) close-up of the initial volume fraction field for a test using an irregular mesh.

Bubble shapes on a polyhedral grid at a sequence of times into the simulation are shown in Figure 8 and magnified around the interfacial region in Figure 9. A qualitative comparison of time histories of the bubble radius modelled using the polyhedral and uniform quadrilateral grids is shown in Figure 10 and it demonstrates that the current method retains on irregular grids approximately the same accuracy attained using uniform grids.

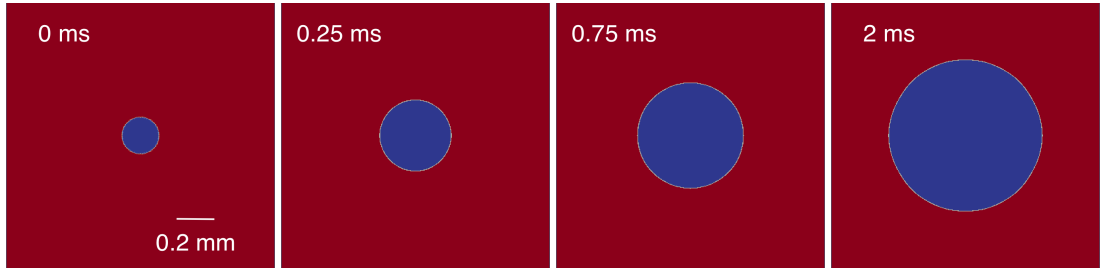

Figure 8

Volume fraction distribution at a sequence of times for bubble growth simulation using a polyhedral mesh.

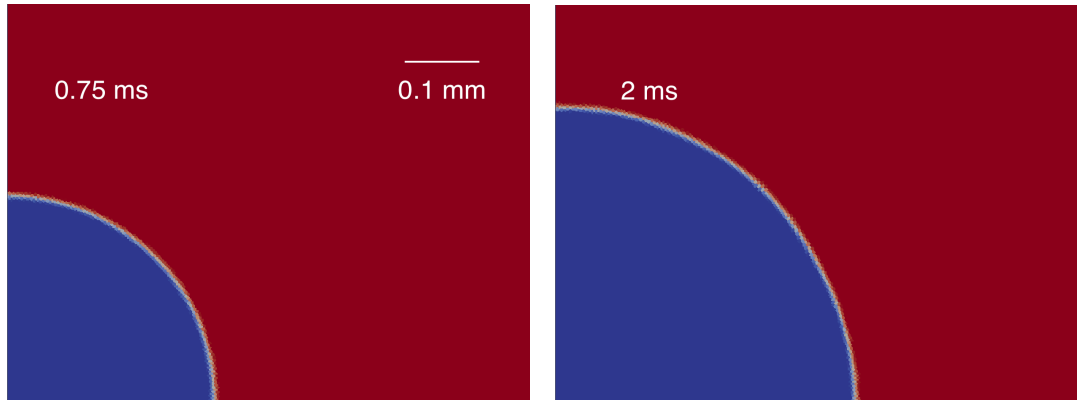

Figure 9

Magnified view of the volume fraction distribution in the interfacial region for bubble growth simulation on a polyhedral mesh.

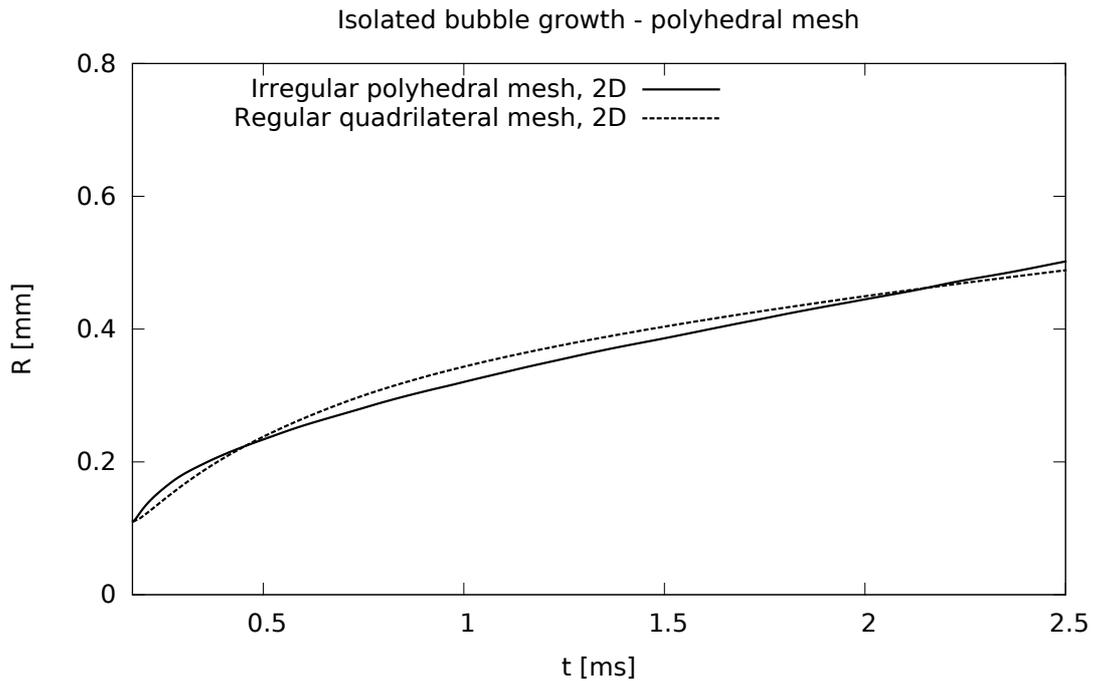

Figure 10

Comparison between time histories of the bubble radius modelled using a uniform translationally symmetric grid ('2D', dashed line) and a non-uniform polyhedral translationally symmetric grid ('2D', solid line).

**5.1.4 Modelled bubble growth vs analytical solution – sodium**

Comparison with analytical solution was repeated for the case of a sodium vapour bubble. Liquid sodium is used alongside water as coolant in nuclear reactors [49] and therefore, as for the case of water, it is important to study its behaviour in conditions when nuclear reactions generate enough heat to cause the fluid to boil; the fluid is also used in single-phase liquid and phase-change receivers of solar power plants [50]. The properties of saturated sodium [51, 52] in conditions typical of its industrial applications are listed in Table 4. The thermal conductivity of liquid sodium is about 70 times higher than liquid water. The test has been conducted considering the same liquid superheat of 3.1K of the water-boiling case and taking the sodium fluid properties from Table 4.

| Properties of saturated sodium at 1.48 bar (1200K) | | |
|---|---|---|
| **Property** | **Vapour** | **Liquid** |
| Dynamic viscosity $\mu$ [$Pa \cdot s$] | $1.8 \times 10^{-11}$ | $152.9 \times 10^{-6}$ |
| Density $\rho$ [$kg/m^3$] | 0.39 | 732.0 |
| Specific heat capacity $c$ [$J/Kg/K$] | 2750.0 | 1250.0 |

| Thermal conductivity $k$ [$W/m/K$] | $48.0 \times 10^{-3}$ | 47.2 |
|---|---|---|
| Surface tension coefficient $\sigma$ [$N/m$] | colspan 0.115 | |
| Latent heat of vaporization $h_{lv}$ [$J/Kg$] | colspan $3840.0 \times 10^3$ | |
| Gas constant $R = R_g/M$ [$J/Kg/K$] | colspan 361.0 | |

Table 4
Properties of saturated sodium at 1.48 bar and 1200K.

Initial conditions of the simulation were computed with Scriven's theory at 4 ms into bubble growth (in this case $\tau \approx 0.2\ ms$), corresponding to an initial radius $R_0$ of 2 mm and initial liquid thermal boundary layer thickness of 0.386 mm. As for the previous case of water, agreement with the analytical solution is observed as the grid is increasingly refined, as shown in Figure 11.

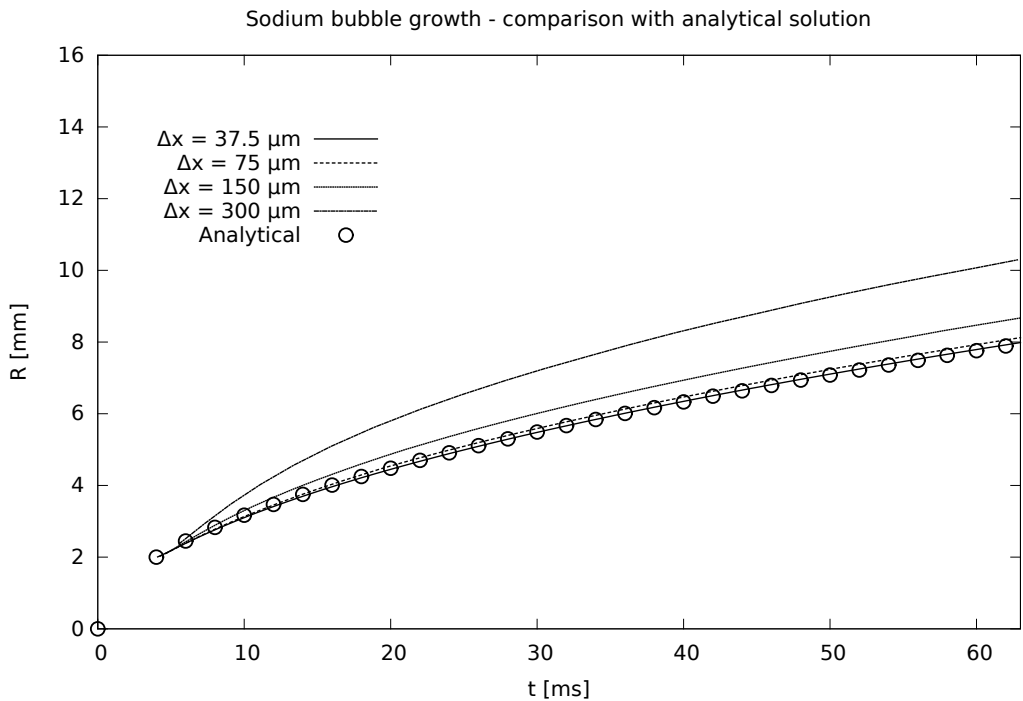

Figure 11

Comparison with analytical solution of modelled time history of bubble radius for the case of bubble growth in isolation in a uniformly superheated pool of liquid sodium.

**5.2 Comparison with experiment**

The experiments of Ref [44] consist of optical observations of steam bubbles growing in isolation in a uniformly superheated pool of water. These are very unusual measurements because normally steam bubbles are observed growing at a heated wall, which introduces the additional complexity of modelling bubble-wall interactions and the very large near wall temperature variations typically found in the liquid around the bubble. On the other hand, the set of experiments of Ref [44] is, to our knowledge, possibly the only one suitable for verification of isolated bubble growth simulations.

Comparison between numerical and experimental bubble growth is shown in Figure 12, showing all three experimental runs from the Ref [44] data set, for the case of 3.1K of liquid superheat, corresponding to a bubble growth constant $\beta = 9.4$. Numerical values of the bubble radius follow closely the experimental data are always well within the bounds of the experimental uncertainty of $\pm 10\%$ estimated in Ref [44].

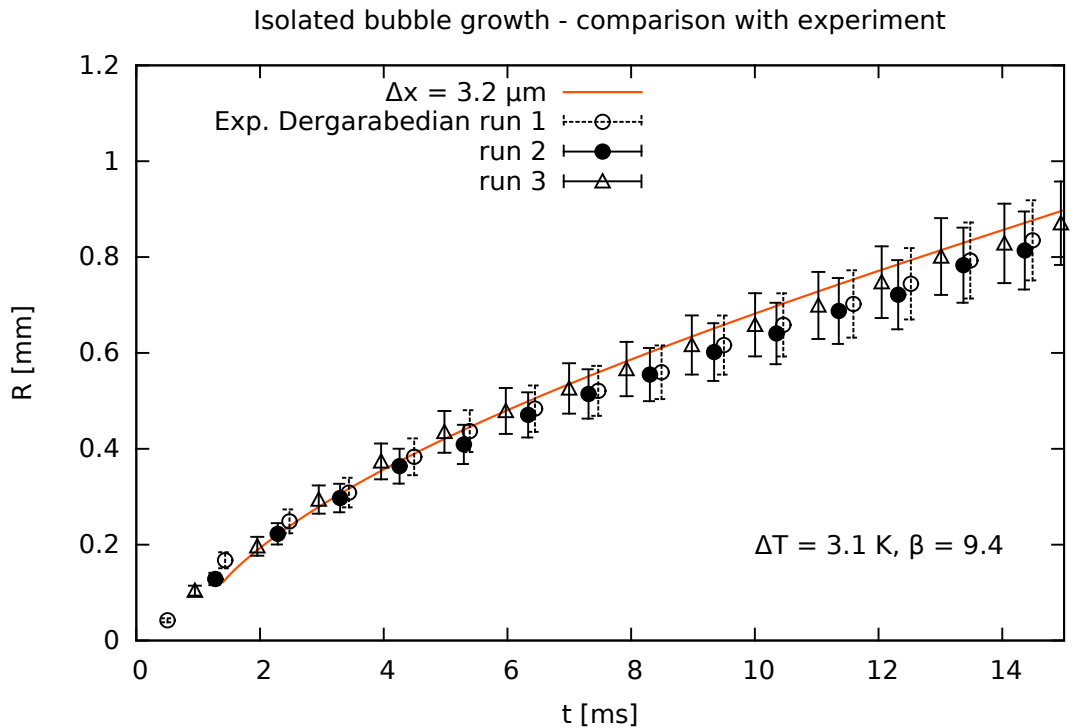

Figure 12

Comparison between numerical (red solid line) and experimental (black/white symbols) bubble growth.

## 6 **RISING STEAM BUBBLE UNDER GRAVITY**

The final validation test considers the simultaneous growth and rise of a steam bubble in an initially stationary pool of superheated water under gravity. Experimental data for comparison are taken from Ref [45], reporting three series of bubble growth observations at similar superheat levels of 3.2, 3.5 and 3.6 K. The axisymmetric simulation domain is a rectangle of 6.0 x 10 mm$^2$ where a spherical bubble is initialized 1.5 mm above the bottom boundary. All boundaries are slip walls except for the symmetry axis and for the outflow boundary at the top of the domain. The bubble is initialized according to the Scriven solution for $\Delta T = 3.5K$ at a time $t_0 = 0.003\ s$, corresponding to an initial radius of 0.473 mm, initial thermal boundary layer thickness of 55 micron and growth factor $\beta = 10.5$. The bubble simultaneously increases in size due to mass transfer and rises in the pool of liquid due to buoyancy, as indicated in Figure 13, showing close-up views of temperature and velocity distributions, for a uniform discretization with $\Delta x = 5.0 \ \mu m$. The fields are displayed at different times until 0.048 s into the simulation i.e., the latest time at which an experimental value for the bubble radius is available. Drag due to relative motion between the rising bubble and the surrounding stationary liquid causes the bubble shape to depart from a sphere, while currents generated by bubble rise under the effect of gravity distort the temperature distribution around the bubble, which loses its spherical symmetry, and generate a wake of cold liquid at positions swept by the bubble's vertical motion. Comparison between modelled and measured bubble growth is indicated in Figure 14. Values of the bubble radius have been divided by the group $2\beta\sqrt{D_l}$ for comparison with the Ref [45] experimental data, which are presented in such a form. The comparison increases confidence in applying the current method to situations where interface motion is not solely due to phase change. External forces such as gravity and hence non-trivial velocity fields causing convective heat transport are captured adequately, at least in conditions of interest and to the extent where available experimental data on bubble growth provide a meaningful comparison.

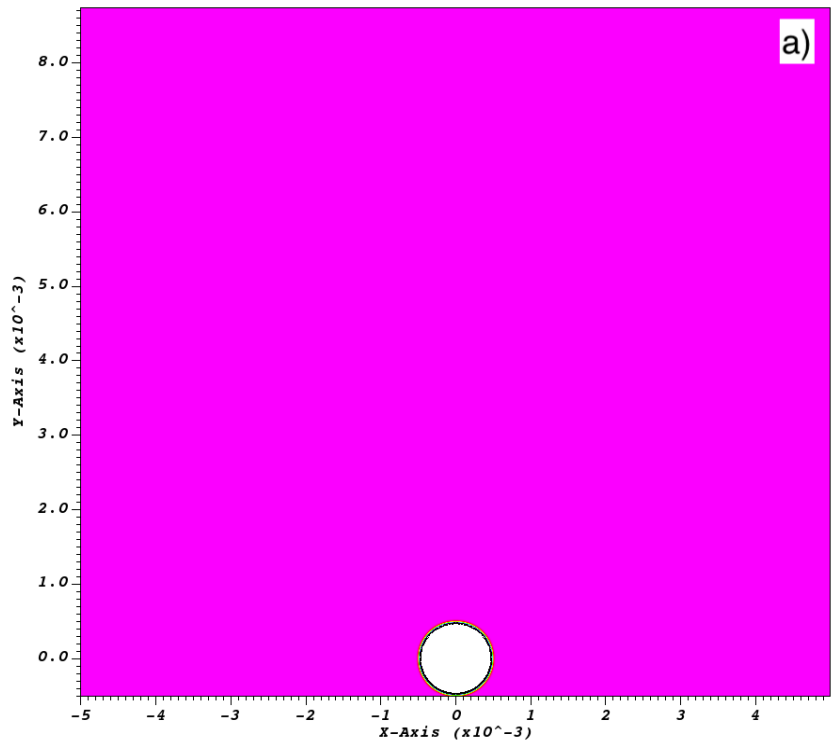

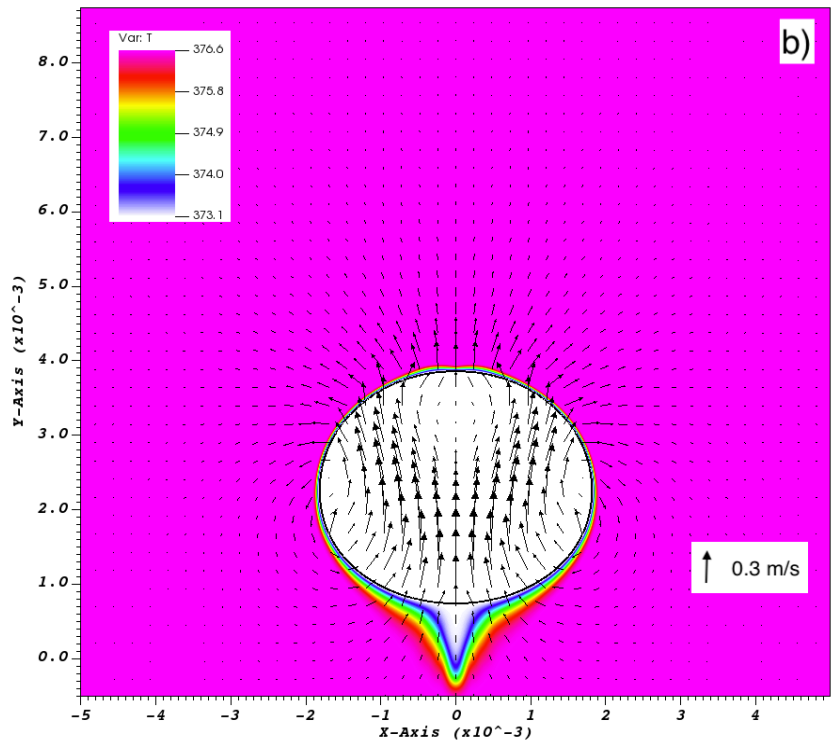
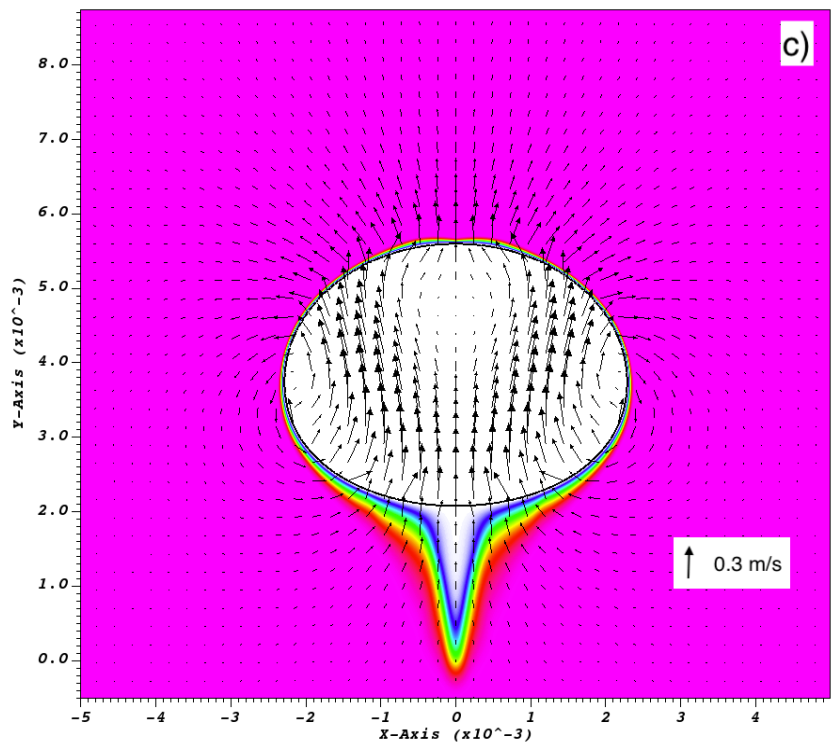

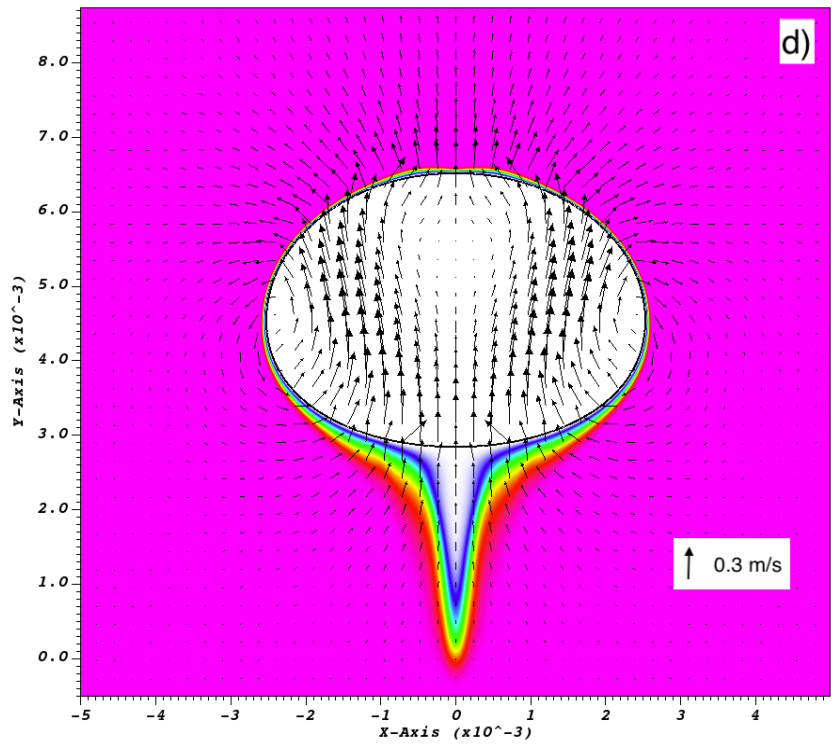

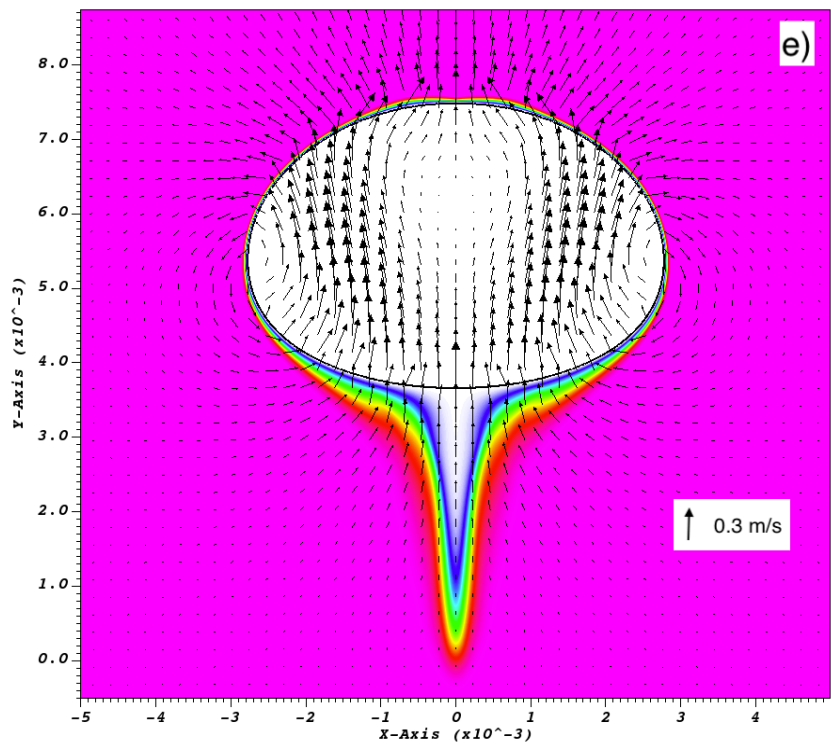

Figure 13
Close-up views of temperature and velocity distributions for the rising bubble test case at 0.003 s (initial condition, panel a), 0.028 (b), 0.038 (c), 0.043 (d) and 0.048 s (e).

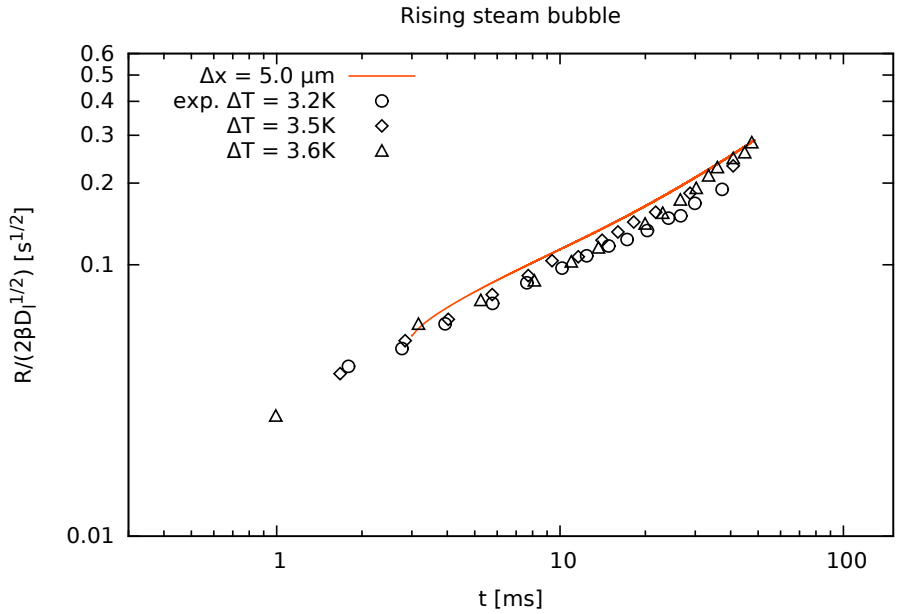

Figure 14
Comparison between modelled and measured bubble radius for the Ref [45] rising steam bubble test case.

## 7 **CONCLUSIONS**

This paper presented development of a numerical method to compute interfacial mass transfer for Interface Capturing simulation of two-phase flow. Implementation of the current methodology using the standard interpolation and solution techniques embodied in the OpenFOAM code enabled its application to arbitrary computational meshes, such as polyhedral meshes here employed that are typical of industrial CFD practice. The accuracy of the method, applicable to arbitrary fluids, was tested via application to evaporation and bubble growth problems representative of the boiling of real fluids, in this case water and sodium. The chosen set of problems represents a severe test for Interface Capturing methodologies due to large density ratios and the presence of strong interfacial evaporation; the proposed framework is however not limited to boiling and may be extended to other phase change phenomena. Comparison with laboratory observations of steam bubbles growing in an extended pool of uniformly superheated water demonstrated that the proposed methodology is applicable to modelling phase-change phenomena in realistic conditions.

## 8 **ACKNOWLEDGEMENTS**

This work was supported by the United Kingdom Engineering and Physical Sciences Research Council (EPSRC) through grant EP/T027061/1.